# On new phenomena of photon from modified double slit experiment

Haisheng Liu
Solution Depot LLC, Hudson, OH 44236, USA
(liuhaisheng@hotmail.com)

**Abstract**

A modified double slit experiment of light was implemented. In the experiment, a spatial shape filter is used to manipulate the shape of cross section of laser beam. When this modified laser beam was shined on the double slit, the intensity distribution of laser beam on double slit is asymmetrical. In this way, the laser light was directed to pass through only one or two slits of double slit in different sections. So the which-way information is predetermined before the photons pass through the slits. At the same time, the visible interference pattern can be observed on a monitor screen after the double slit. A couple of new phenomena had been observed from this experiment. The results of this experiment raise questions about Wave-Particle Duality model of quantum theory, which is the foundation for the Copenhagen explanation that is generally regarded as the principal interpretation of quantum theory. As the observed properties from this experiment cannot be fully explained using the quantum theory, especially the Copenhagen explanation, a new model of photon is proposed. The new model for photon should be also applicable to all other micro entity like electron, atoms, molecules etc., according to L. de Broglie assumption.

Section I. **Introduction**

The Quantum Mechanics is one of the most successes scientific theories in human history. Its application can be extended to a wide range of fields. Despite its success for all practical purposes, fundamental concept of the theory has been debated over one hundred years. Among others, the Copenhagen Explanation is generally considered as orthodox. In the core of the Copenhagen Explanation, there are several basic assumptions including, Schrödinger Equation, Probability Explanation of wave function, Wave-Particle duality, Principle of Complementarity, Uncertainty Principle. Copenhagen Explanation, however, was found insufficient explaining some real world experience [1]. A couple of examples of such deficiency are the superposition principle (Schrödinger cat paradox), and the famous EPR paradox. As results, more than a dozen of alternative explanations had been proposed over years to address these paradoxes, including, GRW, Pilot Wave, De-coherence, Many World, etc. Nonetheless physics community has not fully accepted any of them.

Wave-Particle Duality hypothesizes that any of the micro entity naturally comes with both wave and/or particle properties. The Principle of Complementarity states that the micro entity could display only one of the either at specific event when it is observed or measured, but not both. In other words, it is either wave or particle at a specific time, not wave and particle at same time. In the last several decades, with the progression of the research, it is commonly accepted that it is also possible to observe both wave and particle behavior at the same time, but with limited knowledge about either side.

The origin of the Wave Particle Duality comes from observation of wave like behavior and particle like behavior of light and micro entity of matter in diffraction and interference experiments, especially in the famous double slit experiment. Until today, we still do not fully understand what is governing these micro entities that demonstrate both behaviors in the background even in single electron or photon experiments.

The double slit experiment was a fundamental experiment, which establishes the wave characteristics of light by its interference pattern on the monitor screen, when it is first performed more than two hundreds years ago. Besides light, the double slit experiment was also carried out to explore other matter particles like electron, neutron, and even relative larger molecules during last century. Single photon and single electron double slit experiments were also reported. All of these experiments have firmly confirmed the interference pattern on the monitor screen after a certain time lapse of recording. The interference pattern is commonly considered as a proof of wave characteristics of matter particle. At the same double slit experiment, it is also displays a single spot on the screen, which is showing a typical particle characteristic of both light photon and matter particle. What remains unexplained is how and why the observed interference pattern was formed from a large number of particles in a double slit experiment.

Over the past hundred years, various versions of double slit experiments were performed in an effort to explain interference pattern in the same experiment setup and same time domain, specifically, the relationship between the interference pattern and the which-way information. For the which-way information, most of the experiments focused on the region right before or after the double slit [2], or far away from double slit [3]. All of these different methods were attempted to detect the particles, with minimal disturbance to the particles, at the time the particles passing through the double slit so the which-way information could be obtained. Debates concerning observation of these experiments centered on how much information had been obtained, and/or whether the observed phenomena can be fully explained using current theory. But there is no firm conclusion about it.

In this article, a new modified double slit experiment is presented. The design idea of this experiment is different from previous reports that the which-way information is not detected, but predetermined. With this design, the which-way information is obtained without measurement and disturbances to the photons during the experiment. This experiment yielded new evidence demonstrating weakness of the basic concept of the Wave-Particle duality model. Other assumptions in quantum theory may also need to be updated. It may even reshape our knowledge of micro world if the experiment's conclusion is correct.

As a convention in this paper, left and right slit means from viewer at the position of camera (behind monitor screen). Also the terminologies of bright slit and dark slit refer to the slit was shined or not shined with a laser, respectively (see experiment section below.).

Section II  **Experiment Setup**

In this experiment, the laser beam is manipulated by a beam expander and a spatial shape filter that is used to shape beam cross section of the laser beam, before reaching the double slit. The laser beam pattern at the double slit was controlled so that it could have one slit with laser shined on it (bright slit), but not the other (dark slit). By control the laser beam far away before the double slit, the laser photons are not disturbed when passing through the double slit. In other words, the which-way information is pre-defined. The distinguishability can be estimated by the projection method of the laser beam (see below).

Figure 1 shows the experimental setup that consists of three parts. The first part is the laser beam expander with the beam shape filter. It includes a lens of short focus length, two concave mirrors of the same long focus length, and a beam shape filter between the two concave mirrors. The second part is a pinhole (orifice), a double slit, and a slit blocker. All of these elements are mounted on two-dimensional adjustable micrometers. The pinhole and the double slit are adjustable horizontally and vertically. The beam blocker can be controlled in the horizontal and along beam direction. The third part is a monitor screen and a camera for the interference pattern observation. The monitor screen is located approximately 2 meters away from the double slit.

For the experiment, the laser beam of a He-Ne laser (Uniphase 1125MP) is expanded by a lens (L, focus length ~3.5 cm) and the first concave mirror (FCM, focus length ~ 72 cm). It is aligned that the laser beam will become expanded parallel light beam after the first concave mirror. The diameter of this expanded laser beam is about 3.5 cm. Then a beam shape filter (SF, height ~ 3 cm and width ~ 7 mm) is inserted in this expanded laser beam path to shape the beam cross section. Figure 1 (b) shows the shape of the shape filter. With this shape, when laser beam is projected on the double slit, both top and bottom portion laser beam will shine both slits of the double slit, but only one portion of left (right) of laser beam will be passed through the right (left) slit of double slit in the two middle sections. The slit width of both slits of the double slit is 0.1 mm and the distance between inner edges of the two slits is 0.3 mm. After the shape filter, the second concave mirror (SCM) is used to refocus the laser beam. The image of shape filter is projected on to the double slit as shown in Figure 1 (c). With this shape, the image obtained in the monitor screen (MS) will contain four different parts. The top and bottom parts are same as standard double slit experiment. The two middle parts are formed from laser beam with predetermined which-way information. A pin hole (PH) is used to reduce scattering light, before the laser beam reaching the double slit. A micrometer adjustable block mechanism (SB, Slit Blocker) is placed right after the double slit, as close as possible. This blocker is used to block one of the two slits one at a time. The interference pattern monitor screen (MS) is located about 2 meters away from the double slit. The interference pattern is recorded by Canon digital SLR Rebel XTI (400D) camera (C, Camera).

The experiment was carried out in three different conditions. Under each of these conditions, the images on the MS are recorded for further process. (I) The double slit is not installed. The image on the MS is just the image of the shape filter. This image will be used to estimate the distinguishability. (II) The double slit is installed and both slits of the double slit are opened. This image is main topic of the discussion. (III) The double slit is installed with one of the two slits is blocked. These two images are compared to the condition (II) and also used to calculate the distinguishability.

After the images of interference pattern are obtained, a custom made program is used to extract the intensity information from the image by adding up the pixel values. The intensity is accumulated on certain vertical and horizontal range and extended over the first order diffraction.

Section III  **Results**

Ideally, by design, only left (right) slit will be shined by light (bright slit) for the two middle regions B and C (Figure 1 (c)). The other slit, right (left) (dark slit) should not get any light. But because of scattering and diffraction light from the edges of the shape filter, there is some light get to dark slit even with a pinhole is placed at focus point to reduce this scattering and diffraction light. The light, which reached to the dark slit, will reduce the distinguishability of which-way information. To estimate the effect of the light that pass through the dark slit, an image of the laser beam cross section on the monitor screen with shape filter in position and without double slit is captured. The projection method is used to calculate the intensity of the light, which passes through the dark slit. The figure 2 (a) demonstrates of the method, the (b) is the image on the monitor screen without double slit, and the (d) shows the intensity distribution with two slit position. If we assume the alignment of the center of the image to the center of spacing between slits, the two slits will have to be located on both sides. The following Table 1 shows the results for the double slit with slit spacing (middle opaque part between two slits) 0.3 mm and slit width 0.1 mm. The intensity ratio of the lights, which pass through dark slit and bright slit, is about 90%. The average of square of distinguishability of left and right pattern equals to 0.674

Table 1 Estimated distinguishability by projection method

|  | Max | Min | D | $D^2$ |
|---|---|---|---|---|
| Section C | 960304 | 83247 | 0.840 | 0.706 |
| Section B | 1067261 | 117171 | 0.802 | 0.644 |

The image of interference pattern from the monitor screen with shape filters and double slit is shown in the figure 3. The (a) is the interference pattern with shape filter and double slit in position. As it is designed, the top and bottom section A and D is regular interference pattern observed in every double slit experiment. There is no which way information for them. For the middle two sections B and C, the light is blocked from the center line on the one side by shape filter so the projected image on the double slit will only have light pass through one of the two slits (bright slit). Basically the which-way information is predefined because the asymmetrical light on the double slit. The distinguishability is estimated as discussion above, also calculated below. The interference is also observed for these two sections B and C, but with less contrast, comparing to the section A and D. Also another feature is that the interferece pattern is shift a little bit to the right and left respectively for the right and left slit as bright slit.

The intensity graph of interference pattern of the four sections is shown on the figure 3 (b). Table 2 gives the intensity data from figure 3 and the visibility calculation.

Table 2 Visibility of the two middle sections

|  | Max | Min | V | $V^2$ |
|---|---|---|---|---|
| Section B | 116494 | 40757 | 0.482 | 0.232 |
| Section C | 133446 | 37193 | 0.564 | 0.318 |

From the above visibility data, the average of square of visibility is 0.275

The figure 4 display the diffraction pattern when one of the two slits is blocked. The four sections A, B, C, and D are shown. Now the interference patterns are gone and only the diffraction patterns are displayed for all of the four sections. It is surprised that diffraction patterns are shown in the two middle sections B and C, even when their dark slit are blocked. If the results are compared to the interference pattern obtained from above, without any block of the slit, it is clear that something is blocked so it changes interference pattern to diffraction pattern when the one of slit is blocked. In other words, something has to pass the dark slit to make interference pattern show up even the 90% of light from bright slit correspondently.

It is also noticed that the intensity changes when the dark slit is block. The result is consistent with the calculation from projection method above.

The intensity graph of diffraction pattern of the sections B and C is shown on the figure 4 (c) and (d). Table 3 gives the intensity data from figure 4 and the distinctinguishability calculation.

Table 3 distinctinguishability of the two middle sections

|  | Max | Min | D | $D^2$ |
|---|---|---|---|---|
| Section B | 134563 | 15114 | 0.798 | 0.637 |
| Section C | 141538 | 22432 | 0.726 | 0.527 |

From the above distinctinguishability data, the average of square of distinctinguishability is 0.582. This value is lower than the estimated 0.672, but is in the reasonable range.

From all of above value, the $D^2 + V^2 = 0.857$ to $0.95$. It does satisfy the Englert–Greenberger [4] relationship. So the principle of Completementarity is valid according to this relationship. Even though the current $D^2 + V^2 < 1$, with the improving of the equipment, there is possibility to increase the $D^2 + V^2$ value to greater than one. Nevertheless if the Wave-Particle duality is not valid, there is no base for the Principle of Completementarity.

Section IV **Discussion**
a). Principle of Complementarity

Wave-Particle duality stated that all micro entity should possess wave and particle properties. Here the wave is classical wave and particle is classical particle with integrity and unity. Principle of Complementarity is the rule of how these two properties will be displayed when measured or observed. Basically it is stated that the two properties are mutual exclusive. By that, it means if information of wave characteristics is obtained, then the information particle characteristics cannot be obtained simultaneously, and vice versa. In the case of double slit experiment, it means that if the path information (particle with position information or which slit the light passes through) is known, the interference pattern will be lost. Or if interference pattern were observed, the path information would be lost. The typical example is that if a detector is placed right after one of the two slits, the path information is obtained. But at the same time, the interference pattern is lost. So it is mutual exclusive.

Here from this experiment, the photon of light has predefined path from the shape filter before it pass through double slit. The intensity ratio of dark slit light to the bright slit light is about 10% to 15% from estimated by projection method and measurement from experiment. The visibility of the interference pattern can be calculated from clearly visible interference pattern of the two middle sections B and C. The average value of square of visibility is 0.275 from both sections of B and C. The Principle of Complementarity, specifically the Englert–Greenberger relationship is still satisfied with $D^2 + V^2 = 0.857$ to $0.95$. Even though the relationship is still hold. But a couple observations from the experiment make it truly surprised.

b). It is not wave any more

A couple of new features from this experiment can not be easily associated wave behavior, as it is stated in the Wave-Particle Duality model.

The second feature is the lost of contrast in the B and C sections, comparing to the A and D sections. In the point view of wave, the contrast depends on the double slit structure and intensity ratio from two slits. The A/D and B/C had same slit structure. Only difference is the intensity ratio from two slits. The A and D can get almost 100% contrasts, but B and C get much lower contrast. If consider the ratio for the A to D is 1:1 and B to C or C to B is about 1:10, the contrast of B and C sections should be much lower than the current value of about 50%. This means the wave feature can not explain this result. It means the contrast of interference pattern in the B and C sections is not related with the intensity of average intensity of two slits. It is the single photon behavior, which is consistent with the single photon/electron double experiment.

Another feature observed in the experiment is the interference pattern peak shift to left (right) if the left (right) is bright slit in the middle section B and C. In sections A and D, the photons of interference pattern are clearly from one of the two slits. Even though, we do not know which slit it is come from. But it is certain that the photons from both slit contribute to the light intensity of the interference pattern. This is same as the accumulation of the interference pattern from single photon double slit experiment. For the sections B and C, since the light photons are mainly from one of the two slits (bright slit). Naturally, the sum of the interference pattern from sections B and C would partially construct the full interference pattern of A and D. Or at least, it should have some relationship of them. But when the interference pattern of the two, sections B and C, adds up, it become a big broad peak like diffraction pattern. It is barely related back to the interference pattern from sections A and D.

This point is truly hard to be understood by pure wave property. According to wave theory, the peak of interference is determined by the phase difference of two interference sources. In the case of double slit, the phase difference can be calculated by the distance offset from the central axis and wavelength of the light. In this experiment, the phase difference of all sections, A, B, C, and D, should be same. So the peak position of the interference should be same. The shift of peaks in the section B and C has made wave properties of the photon in vain.

c). Nonlocal and Divisible Photon

The initial idea of this experiment is to answer the question raised from single photon double slit experiment: which slit the photon passes? The standard answer from Copenhagen explanation is that the question is invalid until you observe it. To observe it, a detector is placed right after one of the two slits to observe that the which-way the photon passes, then the interference pattern is lost. But if the physical realty is that photon is particle like as it make click on the target screen, it should always like that. It should not magically change from localized particle to delocalized wave at it own will or experimenter's will. The photon should pass only one slit. Then question is how the interference generated and where is the wave behavior or interference pattern come from.

This experiment does answer above question. Comparing the images from Figure 3 and Figure 4, both figures were obtained with shape filter in place, but with slit blocker in non blocking position (Figure 3) or in position of blocking one of the two slits (Figure 4). For the region A and D, it is easily understandable. The light path in one of two slits was blocked. The which-way information is obtained and the interference pattern changes to diffraction pattern simultaneously. It is exactly as described by Principle of Complementarity. But for the section B and C, the result is difficult to understand. As estimated in the section III, the intensity ratio of light which pass through two slits is about 10%. That means the 90% of light intensity on the monitor screen is from bright slit and 10% from dark slit. When any slit is blocked, either dark slit or bright slit, the interference is disappeared and diffraction pattern does show up for all of 4 sections. Especially considering that the blocked dark slit only contribute about 10% on the monitor screen even when it is not blocked, but the interference pattern is transfer to diffraction pattern. This can not be explained without assuming that "something" has to pass the dark slit. This "something", together with bright photon spot which passes through bright slit, generate interference pattern. This "something" should have the function similar to wave to generate interference pattern. But it is not wave as discussion below. It will be called as Extended Part (EP) in the following discussion. The important view is that photon is constructed with a bright spot in center and EP around it. This is certainly contrary to the popular point of view that photon is local as point particle, or indivisible unit as energy packet or something even it is not point particle.

As it is indicated in the figure 3, the total intensity ratio of the interference fringe to its broad background peak is about 1/1(Right 0.87 and Left 1.32). It is much different from the intensity ratio of dark slit to the bright slit 1/9 to 1/10. So it is hardly to declare the interference fringe peak is from bright slit or dark slit only. it may be from both slits or partial of the bright slit. But light from bright slit definitely should have some contribution to the interference fringe when both slits are opened. Then it is clear that blocking the dark slit does block EP to make interference pattern disappeared.

Comparing the above observation to the single photon double slit experiment, it does have some similarity. At the single photon experiment, photons are evenly distributed on both slits, without any distraction from source to the double slit. That means a photon either pass through left or right slit with full extent of its EP, covered over two slits. But it is only one photon at a time. It is similar as A and D section of this experiment, only it is not single photon experiment for the A and D. The interference pattern is clearly demonstrated at these sections. For the section B and C, the photon reach to the double slit without it full extent of EP because of shape filter. That causes the interference pattern peak shift and lost contrast (see discussion below). To estimate for the pattern changes, it is necessary to have full knowledge of the EP and rule of interaction with bright spot.

With development of ultrafast laser, it is now known that interaction time of photon with material is at scale of femtoseconds. That will translate as about microns in dimension. So it is good estimated the bright spot is about or less than micrometers. From double slit experiment, the extent of the EP should be over millimeters. This is very rough estimate. It does show the size difference of the bright spot and EP.

It had been asked for many years that which slit the photon passed or the photon passed both slits in the single photon double slit experiment. The answer is that the central bright spot of photon does pass through only one slit. But the EP together with bright spot passes both slits. The interaction of this EP with the bright spot after double slit creates the interference pattern.

d). New model of photon

As discussed above, there are two new observations from this experiment: (1) Photon is divisible with its EP and bright spot; (2) The EP surrounding bright spot is not wave. That means Wave-Particle duality model is no longer valid. If the Wave-Particle duality is not correct, there has to be some thing else to explain the interference pattern in the double slit experiment. The particle model itself as in classical point of view can not explain the interference feature of photon. Based on features discussed above of EP, a new model of photon was proposed to explain the experiment results. Even though it may be too simple and bold just based on the current experiment, I am hoping it can stimulate the further research interesting on this field. (1) The photon is not Wave-Particle as what it is assumed in current quantum mechanics. It is constructed by two parts: the bright spot in center and more extended part around it -- EP. (2) The first one is the bright spot, which could be observed by our eye and instruments. It is the bright spot take charge the interaction of photon with its environment. The size of the bright spot part should be small. The second part is more extended, EP, which cannot be observed as light. But it may be observed as we know more about it. (3) The motion of the bright spot is governor by the EP. (4) Some portion of the EP could be separated from bright spot and other EP. (5) The EP will be only existed when the bright spot in motion. This last one shall be true for matter particle too. This EP is not as field idea of electric field or other field. It may have some common properties as field. Certainly it has some more features, like separate able from core bright spot. This is simple model without any mathematical formulas. But it may give us some hint for the future theory. Further details should be discussed in another paper which in process.

With this model in mind, the experiment result is much easy to be understood. In the typical double slit experiment, like the section A and D in this experiment. The photon's bright spot together with dark field pass one slit. Then this dark field interacting with other dark field, passed through other slit, to change the moment of the bright spot of photon. This will create the interference pattern on the target screen. It is also easy to understand the interference pattern in the section B and C. That is why when the one slit is blocked, the interference changes to diffraction pattern. For the feature of contrast and peak shift of sections B and C, it should be explainable by the fact of shape filter. The shape filter does block some light, which will reach to dark slit if not blocked. At same time, the shape filter also disturbed the dark field of the other photon. So when these photon reach to double slit, it accompanied dark field is not symmetrical to the double slit when it get to the double slit. That makes some one can interference and some not. So the peak shifts because of the asymmetrical dark field.

With this model in mind, it is possible to solve other quantum theory paradoxes easily. Wave function collapse is not necessary because the EP will automatically gone when photon hit screen and stopped, just like electric field is gone when electron is discharged by hit a target. Entanglement may also be easily explained by EP interaction with bright spot. Remember, the EP could be separated from bright spot far away. Also, the superposition principle may not need if it is EP. Last one, if it is determinist, no more probability any more. Of course, all of the features of quantum mechanics have to be carefully examined with this model. It could be much more experiment with papers/books to explore all of the possibilities. I do hope this experiment/paper could be stimulating the research of quantum theory further.

Section V   **Conclusion**

A new double slit experiment was carried out. The laser beam image on double slit, predefined by shape filter, was be used to obtain the which-way information. There is nothing to disturb the photons at the moment when they pass through double slit. At the same experiment, the interference pattern is clearly observed on monitor screen even with asymmetrical image on the double slit.  The Principle of Complementarity had been tested by the information obtained. The relationship of Englert–Greenberger duality is verified with $D^2 + V^2$ =0.857 to 0.95. It is much more important that this experiment had proved that photons have to passes both slits to generate interference pattern on the monitor screen. The photon is divisible unit, not as popular point of view that it is indivisible, with the bright spot and Extended Part (EP). The photon is not the classical wave/particle any more. At least, there are two parts.  The one part is the bright spot as we usually observed and the other part is the EP, which is not known before. The interaction of EP governs the motion of the bright spot in free space. Without the EP from other slit, the interference pattern was changed to diffraction pattern magically. Accordingly, a new model of photon is proposed to explain this experimental fact. This model will be certainly applied to all other micro entity of matter like electron, neutron, atoms, and molecules etc. Because of all of these, a theory may be needed to describe this phenomenon and solve the paradox related to quantum theory.

It is certain that this new model will definitely reshape the knowledge of micro entity and related theory if it is true. Our knowledge about this new model and theory is really limited currently. Both experiment and theory work is needed to gain knowledge about micro world.

**Figure Caption:**

Figure 1 Experiment Setup. (a) Sketch of experiment setup. (b) The shape of the Shape Filter. (c) A demonstration image of shape filter on double slit.
HNL – He-Ne Laser; L - lens with short focus length; FCM – First Concave Mirror; SF – Shape Filter; SCM – Second Concave Mirror; PH – Pinhole (Orifice); DS – Double Slit; SB – Slit Blocker; MS – Monitor Screen; C – Camera.

Figure 2 the estimation of distinguishability by projection. (a) The demostration of projection of light. (b) The image of light with shape filter on monitor screen. (c) The full intensity of two middle patterns and projected position of double slit if inserted. (d) The detail central region with double slit position indicated by two horizontal bars.

Figure 3 the Interference pattern with shape filter. (a) The image of interference pattern from double slit with shape filter. (b) The intensity of patterns of every section and sum of sections of B and C. (c) The intensity of patterns of every section in one overlap chart for comparing the B and C shift (see text).

Figure 4 Diffraction pattern with shape filter and blocker. (a) The image of diffraction of double slit with shape filter and block of left slit. (b) The image of diffraction of double slit with shape filter and block of right slit. (c) and (d) The intensity charts of the (a) and (b) for the section B and C.

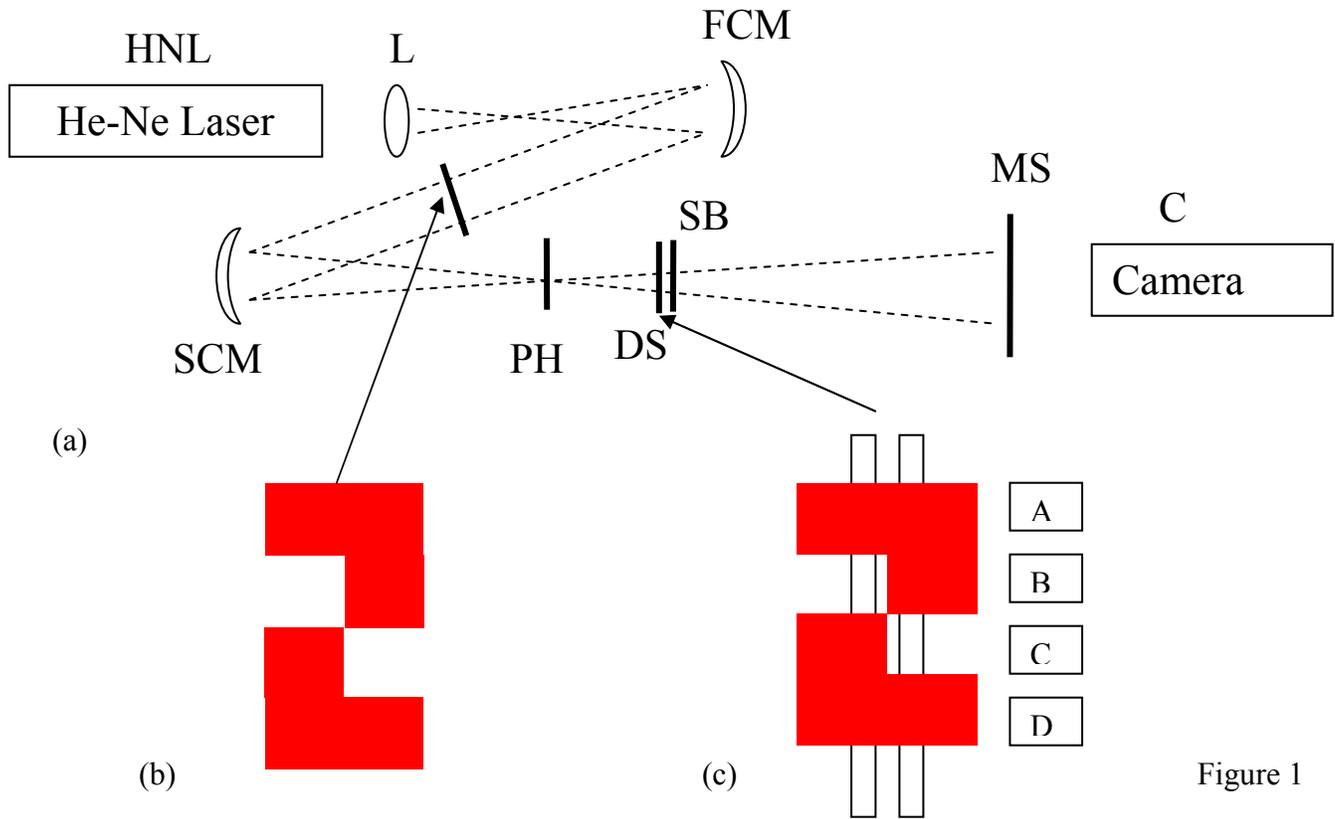

Figure 1

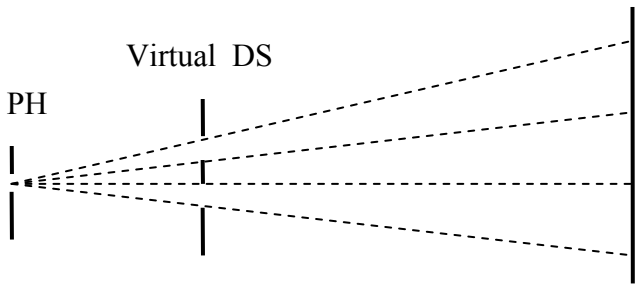

(a)

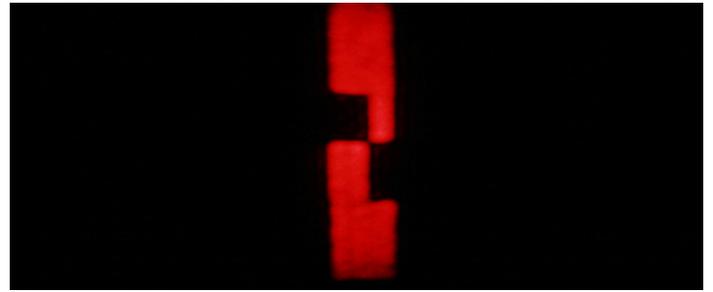

(b)

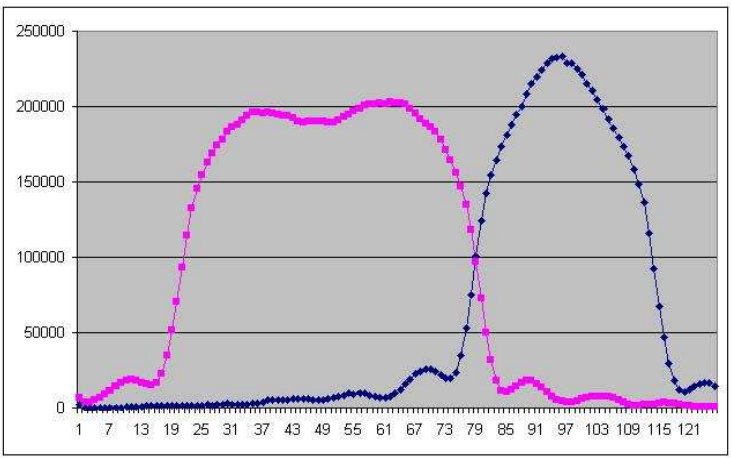

(c)

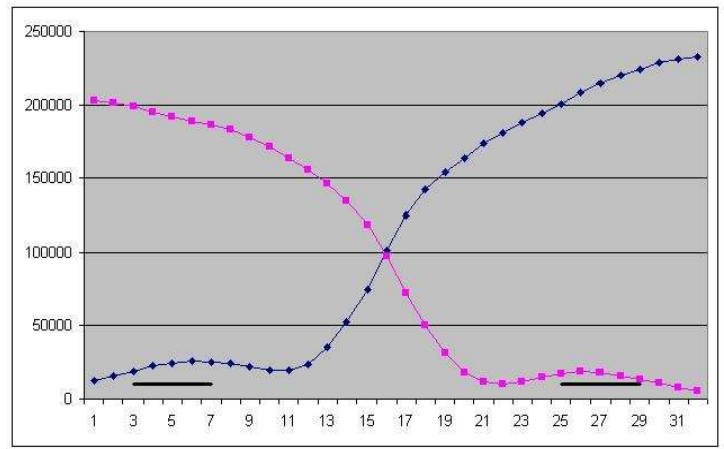

(d)

Figure 2

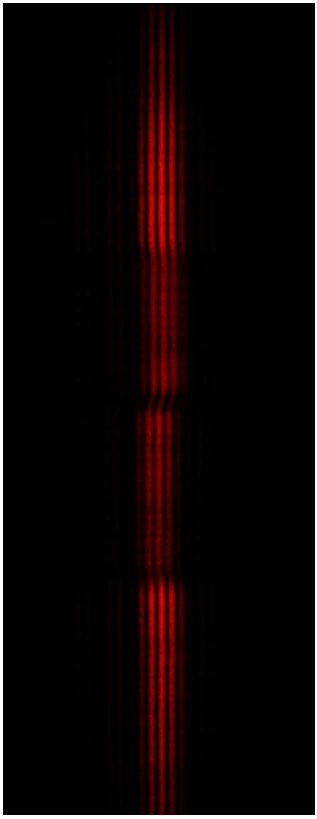 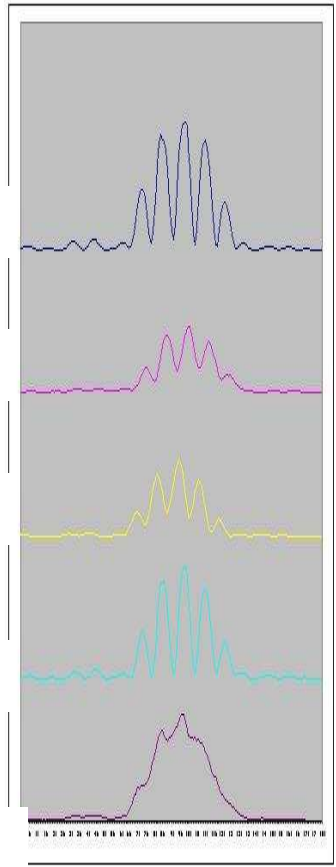 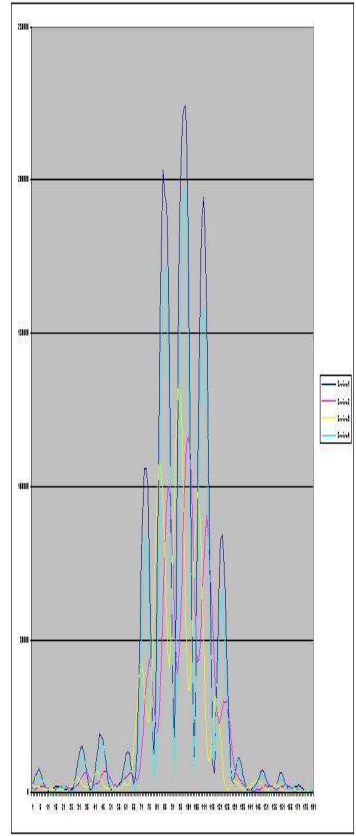

(a)       A   B   C   D   B+C     (b)     (c)    Figure 3

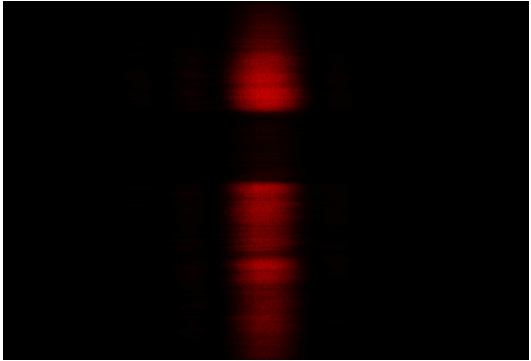 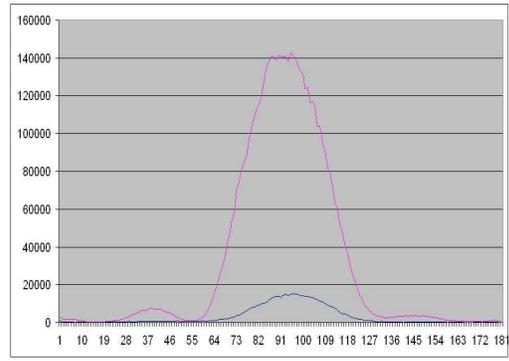

(a)             (c)

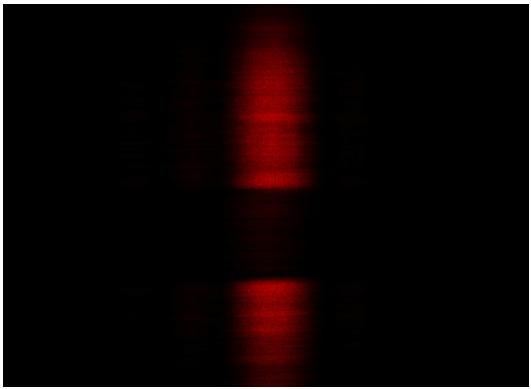 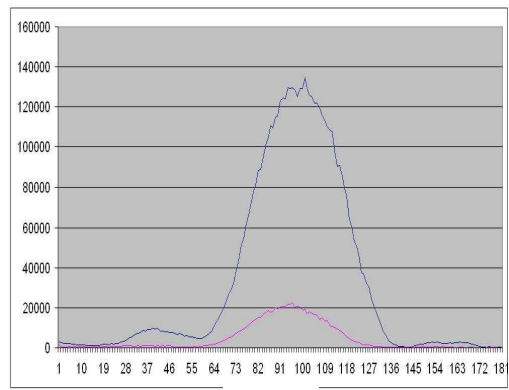

(b)             (d)    Figure 4